\documentclass[aps,prl,reprint,showpacs,superscriptaddress]{revtex4-1}% Physical Review B
\usepackage{graphicx}% Include figure files
\usepackage{latexsym}
\usepackage[T1]{fontenc}
\usepackage{amsmath}
\usepackage{endnotes}
\usepackage{lmodern}
\input{epsf}

\begin{document}

\preprint{APS/123-QED}

\date{\today}

\title{Origin of the tetragonal-to-orthorhombic (nematic) phase transition in FeSe: a combined thermodynamic and NMR study}
%the iron-based superconductor 
\author{A. E. Böhmer}
\affiliation{Institut für Festkörperphysik, Karlsruhe Institute of Technology, 76021 Karlsruhe, Germany}

\author{T. Arai}
\affiliation{Department of Physics, Graduate School of Science, Kyoto University, Kyoto 606-8502, Japan}

\author{F. Hardy}
\affiliation{Institut für Festkörperphysik, Karlsruhe Institute of Technology, 76021 Karlsruhe, Germany}

\author{T. Hattori}
\affiliation{Department of Physics, Graduate School of Science, Kyoto University, Kyoto 606-8502, Japan}

\author{T. Iye}
\affiliation{Department of Physics, Graduate School of Science, Kyoto University, Kyoto 606-8502, Japan}

%\author{P. Schweiss}
%\affiliation{Institut für Festkörperphysik, Karlsruhe Institute for Technology, 76021 Karlsruhe, Germany}

\author{T. Wolf}
\affiliation{Institut für Festkörperphysik, Karlsruhe Institute of Technology, 76021 Karlsruhe, Germany}

\author{H. v. Löhneysen}
\affiliation{Institut für Festkörperphysik, Karlsruhe Institute of Technology, 76021 Karlsruhe, Germany}

\author{K. Ishida}
\affiliation{Department of Physics, Graduate School of Science, Kyoto University, Kyoto 606-8502, Japan}

\author{C. Meingast}
\affiliation{Institut für Festkörperphysik, Karlsruhe Institute of Technology, 76021 Karlsruhe, Germany}

\begin{abstract}

The nature of the tetragonal-to-orthorhombic structural transition at $T_s\approx90$ K in single crystalline FeSe is studied using shear-modulus, heat-capacity, magnetization and NMR measurements.  The transition is shown to be accompanied by a large shear-modulus softening, which is practically identical to that of underdoped Ba(Fe,Co)$_2$As$_2$, suggesting very similar strength of the electron-lattice coupling. On the other hand, a spin-fluctuation contribution to the spin-lattice relaxation rate is only observed below $T_s$. This indicates that the structural, or ``nematic'', phase transition in FeSe is not driven by magnetic fluctuations.

%The nature of the tetragonal-to-orthorhombic structural transition at $T_s\approx90$ K in single crystalline FeSe is studied using shear-modulus, heat-capacity, magnetization and NMR measurements.  The transition is shown to be accompanied by a large shear-modulus softening associated with a large nematic susceptibility. The shear-modulus softening of FeSe is practically identical to that of underdoped Ba(Fe,Co)$_2$As$_2$, suggesting very similar strength of the electron-lattice coupling. On the other hand, a spin-fluctuation contribution to the spin-lattice relaxation rate is only observed below $T_s$. This indicates that nematicity in FeSe is not driven by magnetic fluctuations.

\end{abstract}

%pnictides and chalcogenides, thermodynamic properties, effects of pressure on the phase diagram

\maketitle

One of the most intriguing questions in the study of iron-based superconductors concerns the relation between structure, magnetism and superconductivity \cite{Paglione2010,Johnston2010,Nandi2010,Fernandes2010,Kontani2011,Fernandes2012,Fernandes2013,Nakai2013,Boehmer2013,Fernandes2014}. 
%Stripe-type antiferromagnetic order often occurs at the same or at a slightly lower temperature than the tetragonal-to-orthorhombic structural distortion and the two types of order are closely related by symmetry. 
Stripe-type antiferromagnetic order often occurs at the same or at a slightly lower temperature than the tetragonal-to-orthorhombic structural distortion and the two types of order are closely related by symmetry. They break the four-fold rotational symmetry of the high-temperature phase, which can be associated with a nematic degree of freedom \cite{Fernandes2010,Fernandes2012}. Superconductivity typically is strongest around the point where the structural transition ($T_s$) and the antiferromagnetic transition ($T_N$) are suppressed by pressure or chemical substitution. Whether the magnetic or the structural instability is the primary one, is still under intense debate \cite{Fernandes2014}, also because of its relevance to the pairing mechanism \cite{Kontani2011,Fernandes2012}. Recently, scaling relations between the shear modulus related to the structural distortion, $C_{66}$, and the spin-lattice relaxation time $T_1$ as a measure of the strength of spin fluctuations, have been proposed \cite{Fernandes2013,Nakai2013} in order to address the above question. They were found to be well satisfied in the Ba(Fe,Co)$_2$As$_2$ system \cite{Fernandes2013,Nakai2013}, where $T_s$ and $T_N$ are in close proximity to each other, suggesting a magnetically-driven structural transition \cite{Fernandes2013}. Clearly, it is of great interest to see if a relation between shear modulus and spin fluctuations is universally observed in other iron-based materials. 

FeSe is structurally the simplest iron-based superconductor and has attracted a lot of attention because of a nearly four-fold increase of its $T_c\approx 8$ K under pressure \cite{Mizuguchi2008}.
%,Medvedev2009,Margadonna2009,Garbarino2009}. 
Moreover, this system is particularly interesting with respect to the relation of structure and magnetism, since it undergoes a tetragonal-to-orthorhombic structural phase transition at $T_s\sim90$ K, similar to that found in the 1111- and 122-type parent compounds \cite{Johnston2010}, but does not order magnetically at ambient pressure \cite{Bendele2010,McQueen2009}. Spin fluctuations at low temperatures were, however, observed in nuclear magnetic resonance (NMR) measurements \cite{Imai2009}. 
Surprisingly, the orthorhombic distortion of FeSe is not reduced upon entering the superconducting state \cite{Boehmer2013} in strong contrast to underdoped BaFe$_2$As$_2$ \cite{Nandi2010,Wiesenmayer2011}, indicating different couplings between structure and superconductivity. This strongly motivates further study of the interplay of structure, magnetism and superconductivity in FeSe. 

\begin{figure}[tb]
\includegraphics[width=8.6cm]{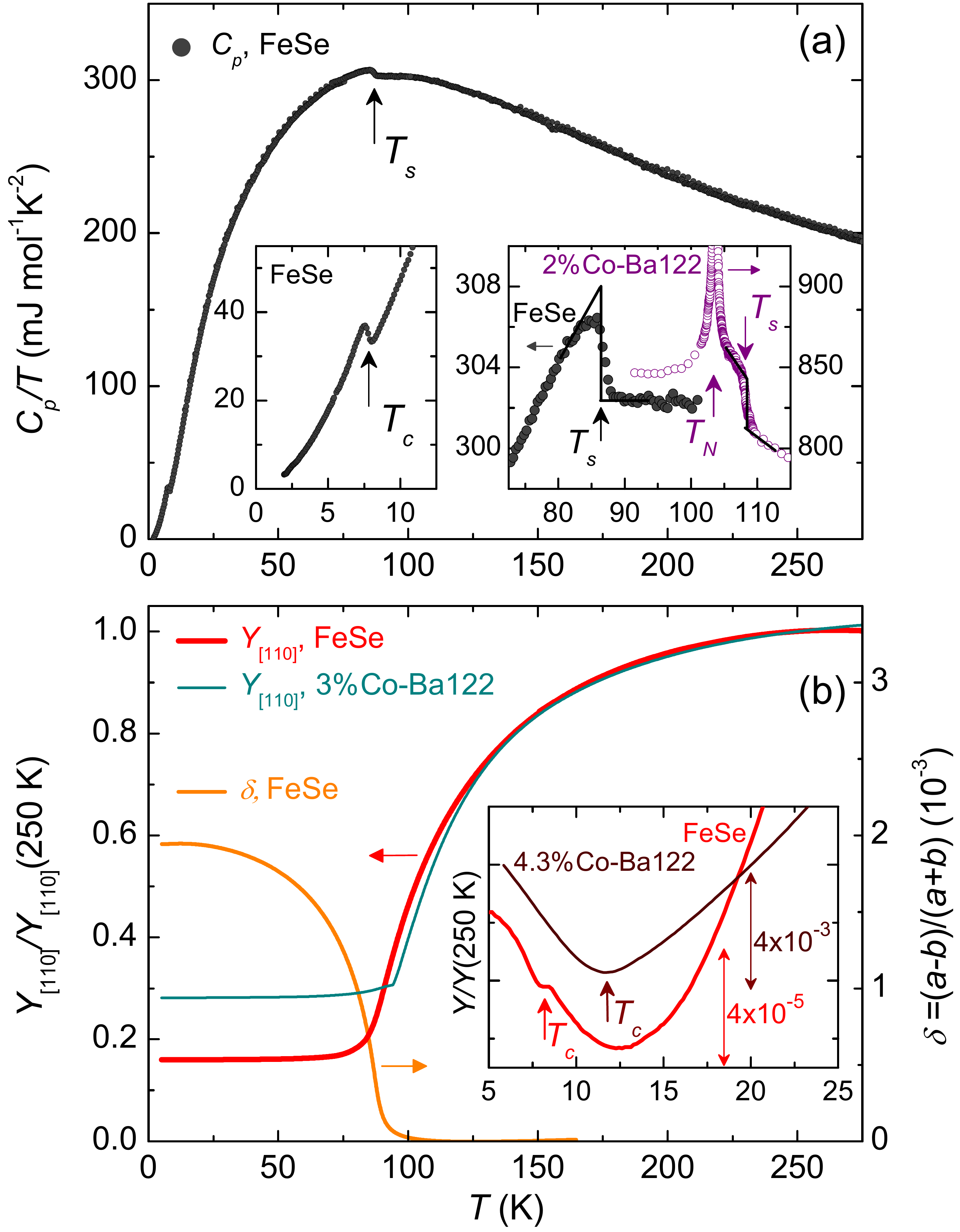}
\caption{(color online) (a) Specific heat $C_p$ divided by $T$ vs. $T$ of an FeSe single crystal. Insets show data around $T_c$ and $T_s$ on an enlarged scale revealing relatively sharp, mean-field-like transitions. Data for 2\% Co-doped BaFe$_2$As$_2$ are shown for comparison. (b) Young's modulus $Y$ (left scale) and orthorhombic distortion $\delta$ (right scale) of FeSe single crystals vs. $T$, compared with slightly underdoped Ba(Fe,Co)$_2$As$_2$. The inset shows data around $T_c$ of FeSe and 4.3\% Co-doped BaFe$_2$As$_2$. Note the different vertical scales in the inset.}
\label{fig:1}
\end{figure}

In this letter, we study FeSe using shear-modulus, specific-heat, magnetization and NMR measurements in vapor-grown \cite{Boehmer2013} single crystals and compare our results to those of underdoped Ba(Fe,Co)$_2$As$_2$. We find that the magnetic fluctuations observed in the NMR data cannot be the driving force for the structural transition, since they set in only below $T_s$. Further, the shear-modulus softening above $T_s$ is found to be nearly identical in FeSe and underdoped Ba(Fe,Co)$_2$As$_2$, possibly suggesting a common origin of the structural transition in both systems. 
%This naturally puts magnetic fluctuations as the origin of the structural transition in the BaFe$_2$As$_2$-based systems also into question. 

Figure \ref{fig:1} shows thermodynamic data of FeSe and, for comparison, of lightly Co-substituted BaFe$_2$As$_2$. A clear mean-field-like anomaly with $\Delta C_p/T_s\approx 5.5$ mJ mol$^{-1}$K$^{-2}$ is observed at $T_s=87$ K in the specific heat of FeSe. The discontinuity is similar in magnitude to the low-temperature Sommerfeld coefficient $\gamma_L=5.7$ mJ mol$^{-1}$K$^{-2}$, suggesting an electronic instability consistent with a recently observed reconstruction of the Fermi surface at $T_s$ \cite{Nakayama2014,Huynh2014,Shimojima2014}. A similar step-like specific-heat anomaly is also seen at $T_s$ of Ba(Fe$_{0.98}$Co$_{0.02}$)$_2$As$_2$ (see inset in Fig. \ref{fig:1}(a)), where $T_s$ is well separated from $T_N$.
The temperature dependence of the orthorhombic distortion $\delta=(a-b)/(a+b)$, derived from thermal-expansion data  \cite{Boehmer2013}, (Fig. \ref{fig:1}(b), $a$ and $b$ are the in-plane lattice constants of the orthorhombic unit cell) also provides a clear indication of the structural transition and is very similar to that of BaFe$_2$As$_2$.

Shear-modulus measurements offer another powerful method for studying the structural transition \cite{Fernandes2010,Goto2011,Yoshizawa2012,Boehmer2014}. If there is an electronic origin of the tetragonal-to-orthorhombic transition, it can phenomenologically be ascribed to the divergence of the susceptibility $\chi_\varphi$ of an electronic and, by symmetry, nematic order parameter $\varphi$, irrespective of its microscopic nature \cite{Chu2012}. In this case, the Landau-type free energy is written as
\begin{equation}
F=\frac{1}{2}\left(\chi_\varphi\right)^{-1}\varphi^2+\frac{B}{4}\varphi^4+\frac{C_{66,0}}{2}\delta^2-\lambda\varphi\delta,
\end{equation}
with bilinear coupling, $\lambda$, between $\varphi$ and the orthorhombic distortion $\delta$, allowed by symmetry, and a bare shear modulus $C_{66,0}$. In consequence, the effective elastic shear modulus, given by 
\begin{equation}
{C_{66}}=\frac{d^2F}{d\delta^2}=C_{66,0}-\lambda^2\chi_\varphi, \label{eq:5}
\end{equation}
is determined solely by the ``phenomenological'' nematic susceptibility $\chi_\varphi$ and the coupling constant $\lambda$ \cite{Salje1990, Fernandes2010, Cano2010, Boehmer2014}.  

In Fig. \ref{fig:1}(b), we show the Young modulus along the tetragonal [110] direction, $Y_{[110]}$, whose temperature dependence was previously shown to be dominated by $C_{66}$ \cite{Boehmer2014}, of FeSe and Ba(Fe$_{0.97}$Co$_{0.03}$)$_2$As$_2$, as measured in a three-point bending setup in a capacitance dilatometer \cite{Boehmer2014}. The significant softening on approaching $T_s$ from above is characteristic of the elastic soft mode, i.e., $C_{66}$ \cite{Yoshizawa2012II,Zvyagina2013}. Strikingly, this softening is practically identical in the two systems, which shows that $\lambda^2\chi_\varphi/C_{66,0}$ of FeSe is practically identical to that of Ba(Fe$_{0.97}$Co$_{0.03}$)$_2$As$_2$, implying that the coupling between nematic order parameter and lattice $\lambda^2/C_{66,0}$ has nearly the same value in the two systems. 

Below $T_s$, the Young modulus is nearly constant and does not show the increase expected for a second-order phase transition, presumably due to the formation of structural twins within the orthorhombic phase \cite{Schranz2012}. Nevertheless, small anomalies around $T_c$ can still be resolved (see inset in Fig. \ref{fig:1}(b)). $Y_{[110]}$ of Ba(Fe$_{0.957}$Co$_{0.043}$)$_2$As$_2$ ($T_c=12$ K) hardens anomalously by $\sim4\times10^{-3}$ below $T_c$, an effect previously observed in overdoped Ba(Fe,Co)$_2$As$_2$ \cite{Fernandes2010,Yoshizawa2012} and interpreted as a consequence of the competition between magnetic fluctuations and superconductivity in the spin-nematic scenario \cite{Fernandes2010}. In strong contrast, $Y_{[110]}$ of FeSe only shows a small step-like softening by $\Delta Y_{[110]}\approx8.5\times10^{-6}$ at $T_c$. The step-like softening is the normal behavior expected at a superconducting transition and is related to the uniaxial pressure derivative of $T_c$ and the specific-heat anomaly via a thermodynamic relation \cite{noteFeSe}. Importantly, the absence of any anomalous hardening of $Y_{[110]}$ related to $T_c$ demonstrates again \cite{Boehmer2013} that the orthorhombic phase and superconductivity do not compete in FeSe, as they do in substituted BaFe$_2$As$_2$ \cite{Fernandes2010,Yoshizawa2012,Boehmer2014}.  
We note that $Y_{[110]}$ hardens slightly by $\sim5\times10^{-5}$ below $\sim12.5\textnormal{ K}>T_c$, which correlates well with the anomalous thermal expansion below roughly the same temperature \cite{Boehmer2013}.

\begin{figure}
\includegraphics[width=8.6cm]{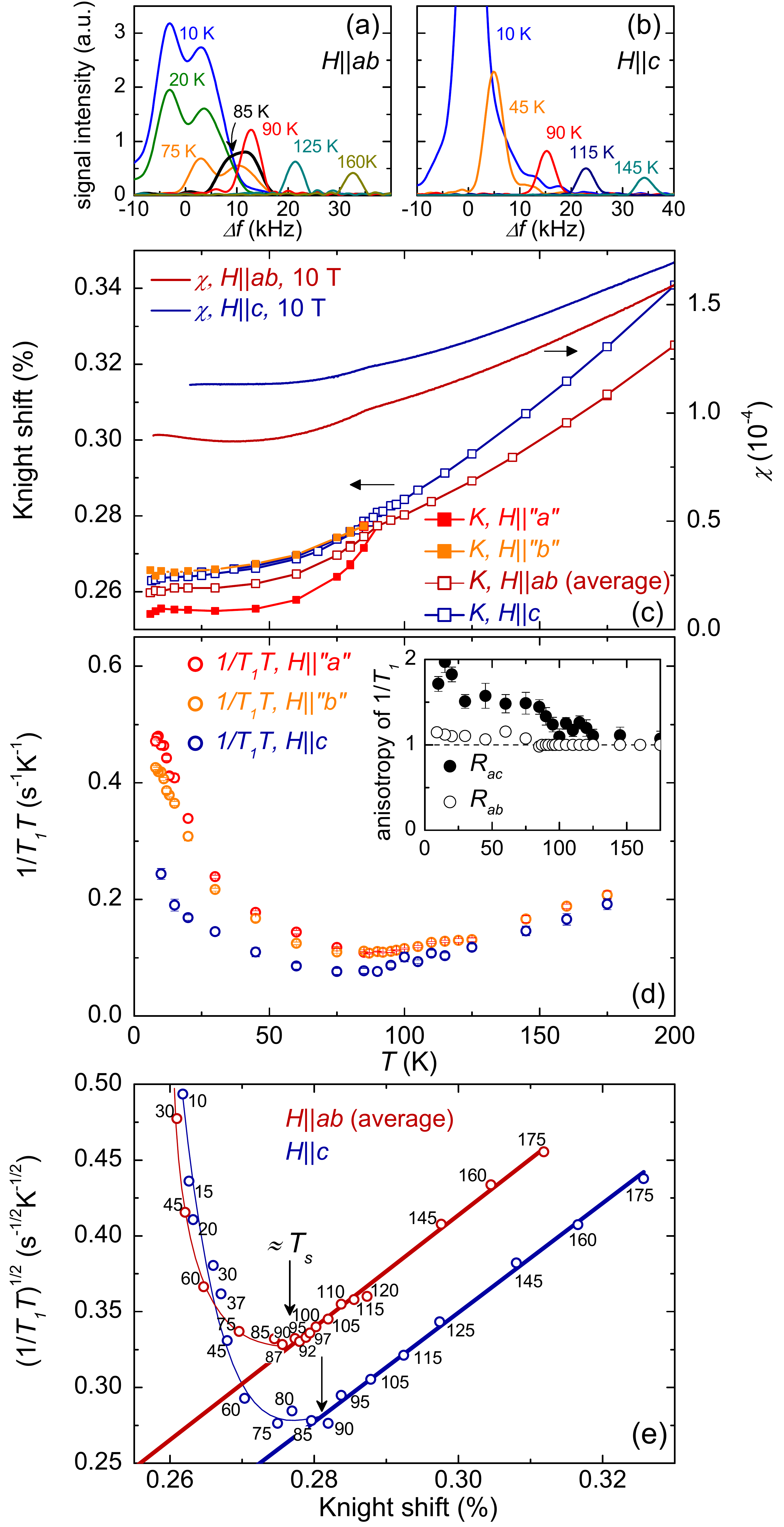}
\caption{(color online) (a),(b) NMR spectra of a collection of $\sim10$ FeSe single crystals with field $H||ab$ and $H||c$, respectively, at $H=9$ T. (c) Knight shift (left scale) and uniform magnetic susceptibility (right scale) for the indicated field directions. (d) $1/T_1T$ and its anisotropy with respect to the applied field $R_{ac}=\frac{\left[\left(1/T_1\right)_{H||\textnormal{``$a$''}}+\left(1/T_1\right)_{H||\textnormal{``$b$''}}\right]/2}{\left(1/T_1\right)_{H||c}}$ and $R_{ab}=\frac{\left(1/T_1\right)_{H||\textnormal{``$a$''}}}{\left(1/T_1\right)_{H||\textnormal{``$b$''}}}$ (inset). (e) Square root of $1/T_1T$ vs. Knight shift with temperature as implicit parameter, indicated in units of K. Bold straight lines show a linear fit to the data for $T>T_s$, eq. \ref{eq:Korringa}, deviations from which demonstrate the emergence of spin fluctuations. Thin lines are a guide to the eye. $K(T_s)$ is indicated by vertical arrows.}
\label{fig:2}
\end{figure}

In order to investigate the microscopic physics, we have performed $^{77}$Se NMR measurements on a collection of $\sim10$ single crystals, aligned by eye, in a field of 9 T. $^{77}$Se has a nuclear spin of $I=1/2$ and therefore no quadrupolar interactions. The resonance lines in the field-swept NMR spectra, observed at a fixed frequency of $f=73.28$ MHz, are very narrow with FWHM of only $5-8$ kHz (Fig. \ref{fig:2} (a),(b)). Interestingly, the resonance lines split below $T_s\sim87-90$ K under an in-plane field, which was confirmed by measurements on only one single crystal. Since NMR is a local probe, this clearly shows the existence of two types of domains having different Knight shifts $K$ in which the field is aligned parallel to either the orthorhombic $a$ axis or the $b$ axis. We arbitrarily assign the smaller Knight shift to domains with $H||a$ (``$a$'') and the larger Knight shift to domains with $H||b$ (``$b$''). Note that a similar observation was reported for LaFeAsO, but was attributed to quadrupolar effects \cite{Fu2012}.

Fig. \ref{fig:2} (c) shows $K$ and the uniform magnetic susceptibility $\chi$, measured in a vibrating sample magnetometer at 10 T. The relatively strong temperature dependence of $K$ and $\chi$ is presumably due to the small Fermi-surface pockets found in FeSe \cite{Maletz2014,Shimojima2014,Terashima2014}. In general, $K$ is given by $K=K_{spin}+K_{chem}$ with $K_{spin}=A_{hf}\chi/N_A\mu_B$ and a temperature independent chemical shift $K_{chem}$. $A_{hf}$ is the relevant component of the hyperfine coupling tensor. Scaling of $K$ and $\chi$ at $T>T_s$ yields $A_{hf}^{aa}=2.49(1)$ $\mu_\textnormal{B}$/T and $A_{hf}^{cc}=3.72(3)$ $\mu_\textnormal{B}$/T. 

Fig. \ref{fig:2} (d) shows the spin-lattice relaxation rate divided by $T$, $1/T_1T$ and Fig. \ref{fig:2} (e) an analysis of its temperature dependence. The data agree qualitatively well with the early data by Imai et al. on polycrystalline samples \cite{Imai2009}. Here, we study in particular the region around $T_s$ and the magnetic-field anisotropy of $1/T_1$. 
In general, there are several contributions to $1/T_1$. For a Fermi liquid, the hyperfine coupling between nuclear spins and conduction electrons results in the Korringa contribution, following the relation
\begin{equation}
{\left(\frac{1}{T_1T}\right)_{\mathrm{FL}}}\propto K^2_{spin}.\label{eq:Korringa}
\end{equation}
Fluctuating transverse magnetic fields provide an additional relaxation process leading to
\begin{equation}
\left(\frac{1}{T_1T}\right)_{\mathrm{sf}}\propto\lim_{\omega\rightarrow0}\sum_\mathbf{q}{F^2(\mathbf{q})\frac{\mathrm{Im}\chi(\mathbf{q},\omega)}{\omega}},
\end{equation}
Here, $F^2(\mathbf{q})$ is a wave-vector dependent form factor and $\chi(\mathbf{q},\omega)$ the dynamic spin susceptibility. The two contributions add up to the total relaxation rate $1/T_1T=(1/T_1T)_{\mathrm{FL}}+(1/T_1T)_{\mathrm{sf}}$.  In order to discriminate between these two contributions, we show in Fig. \ref{fig:2}(e) $\sqrt{1/T_1T}$ plotted versus the Knight shift with temperature as an implicit parameter. From eq. \ref{eq:Korringa}, one would expect the data to fall on a straight line, which indeed holds for $T>T_s$. Importantly, deviations from the Korringa behavior, which signal the emergence of significant magnetic fluctuations, occur only below $T_s$. 

Information about the nature of the magnetic fluctuations may be obtained from the field anisotropy of $1/T_1T$ within the orthorhombic phase. Namely, the $ab$-anisotropy ratio
\begin{equation}
R_{ab}=\frac{\left(1/T_1\right)_{H||\textnormal{``$a$''}}}{\left(1/T_1\right)_{H||\textnormal{``$b$''}}}\approx 1.1-1.2\quad(T<T_s)
\end{equation}
is found to be quite small and nearly temperature independent. 
The ratio of the in-plane average of $1/T_1$ and its $c$-axis value
\begin{equation} R_{ac}=\frac{\left[\left(1/T_1\right)_{H||\textnormal{``$a$''}}+\left(1/T_1\right)_{H||\textnormal{``$b$''}}\right]/2}{\left(1/T_1\right)_{H||c}},
\end{equation}
is $\sim1.5-2$ at low $T$ where spin fluctuations dominate $1/T_1$.
These results are in strong contrast to LaFeAsO, where $R_{ab}$ nearly doubles \cite{Fu2012} and $R_{ac}$ increases strongly from $\approx1.5$ to $\approx3$ \cite{Nakai2012} on decreasing $T$ between $T_s$ and $T_N$. The former observation has been taken as a characteristic of the spin-nematic state, in which spin fluctuations are at the origin of $ab$ anisotropy \cite{Fu2012}. The small value of $R_{ab}$ in FeSe hence suggests the absence of a spin-nematic state at low $T$ in FeSe. 

\begin{figure}[tb]
\includegraphics[width=8.6cm]{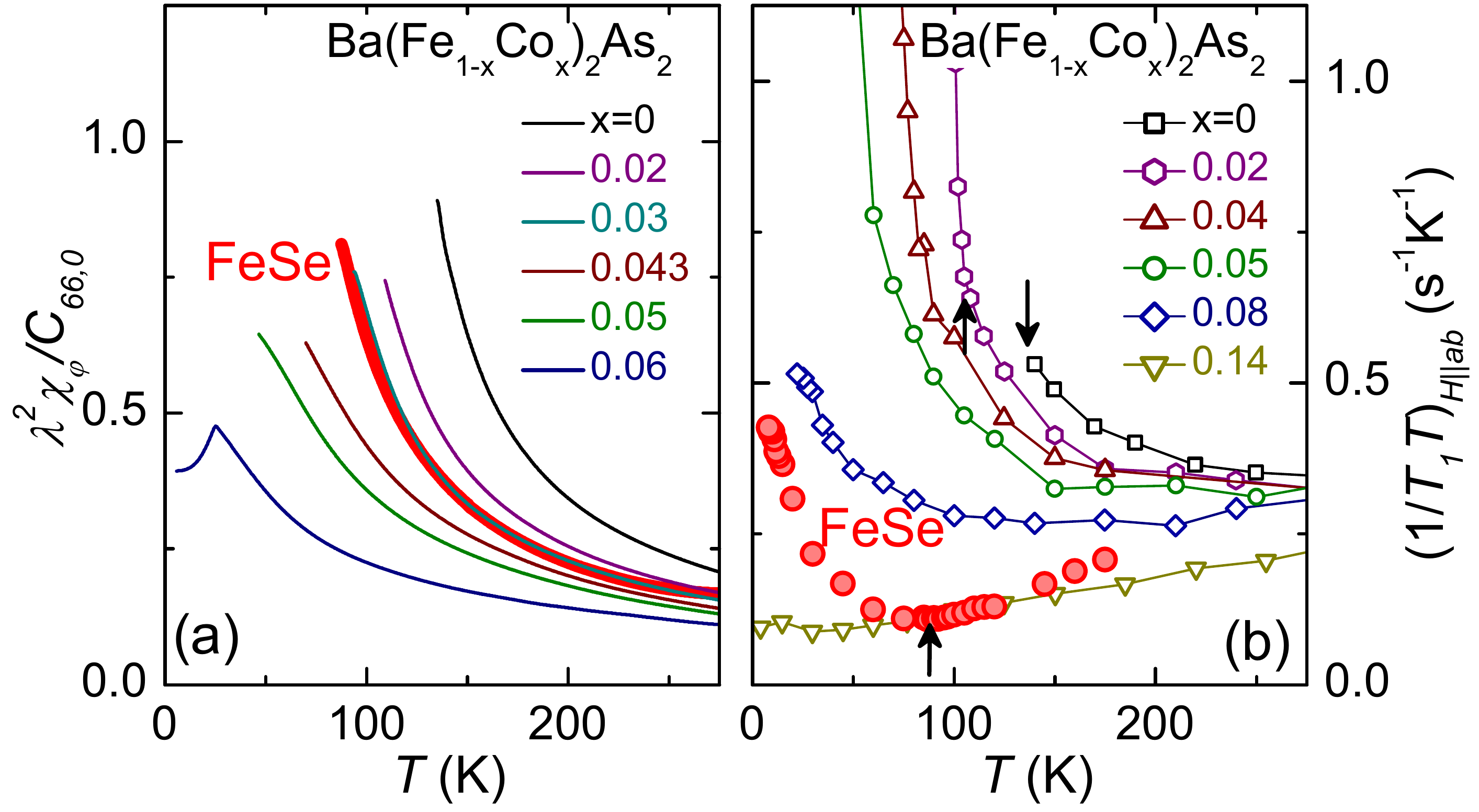}
\caption{(color online) (a) Normalized nematic susceptibility $\lambda^2\chi_\varphi/C_{66,0}=1-C_{66}/C_{66,0}$ of FeSe and Ba(Fe,Co)$_2$As$_2$ (from Ref. \cite{Boehmer2014}). Data on FeSe are found to be practically identical to Ba(Fe,Co)$_2$As$_2$ with the same $T_s$. (b) $1/T_1T$ of FeSe and Ba(Fe,Co)$_2$As$_2$ (from Ref. \cite{Ning2010,Ning2014}) for in-plane field, demonstrating very distinct behavior in the two systems. Arrows mark $T_s$ of BaFe$_2$As$_2$, Ba(Fe$_{0.98}$Co$_{0.02}$)$_2$As$_2$ and FeSe.}
\label{fig:3}
\end{figure}

Fig. \ref{fig:3} summarizes our results concerning the phenomenological nematic susceptibility $\chi_\varphi$, derived from the Young-modulus data \cite{notenemsus} and $1/T_1T$, which are closely related in the spin-nematic scenario \cite{Fernandes2012,Fernandes2013}. 
Remarkably, $\lambda^2\chi_\varphi/C_{66,0}$ of FeSe fits very well into the Ba(Fe,Co)$_2$As$_2$ series (Fig. \ref{fig:3}(a)), showing that the nematic susceptibility and the electron-lattice coupling are very similar, as already argued above. The temperature dependence of $1/T_1T$ of FeSe, on the other hand, clearly does not fit into the Ba(Fe,Co)$_2$As$_2$ series (Fig. \ref{fig:3}(b)). In particular, the large spin-fluctuation contribution to $1/T_1T$, observed up to room temperature in lightly doped Ba(Fe,Co)$_2$As$_2$ \cite{Ning2010}, is not found in the FeSe data. 
%The absence of magnetic fluctuations above $T_s$ implies that they cannot be the cause for the shear-modulus softening and, thus, the structural transition of FeSe. 
Our results for FeSe therefore put the spin-nematic scenario, in which the lattice softening is the result of increased spin fluctuations \cite{Fernandes2010,Fernandes2012,Fernandes2013}, into question, even if the scaling of $C_{66}$ and $T_1$ of Ref. \cite{Fernandes2013} needs not to be strictly valid for non-finite $T_s$ and $T_N$ \cite{Fernandes2013}.

The NMR relaxation data show that the onset of magnetic fluctuations coincides approximately with $T_s$ and that FeSe appears to be close to a magnetic instability at low temperatures. The result seems to suggest that the structural transition triggers the emergence of magnetism. This, however, does not hold under hydrostatic pressure, where spin fluctuations are enhanced \cite{Imai2009}, while $T_s$ is rapidly suppressed \cite{Miyoshi2014}. Possibly, FeSe tends to a tetragonal-type magnetic order, which naturally would not couple strongly to the orthorhombic distortion, as is also suggested by the magnetic-field anisotropy of $1/T_1T$. A magnetic state within a quasi-tetragonal structure has, for example, been observed in Na-substituted BaFe$_2$As$_2$ \cite{Avci2014,Wasser2014}. 

In summary, we have shown that FeSe exhibits a surprisingly similar shear-modulus softening as found in the 122 compounds, suggesting a common origin of the structural transition in these systems. Spin fluctuations only emerge below $T_s$ in FeSe and are therefore argued not to be the driving force of its structural transition. This leaves orbital ordering as a possible driving force and, in fact, ARPES measurements \cite{Nakayama2014,Shimojima2014} find evidence for the orbital ordering scenario. Namely, a strong orbital anisotropy, which is greater than expected from the small structural distortion $\delta$ alone, is observed below $T_s$. 
Finally, our results naturally raise the question of the origin of superconductivity in FeSe, since both orbital and magnetic fluctuations have been considered as a pairing glue for superconductivity in the iron-based materials. 
If superconductivity were mediated by orbital fluctuations, one might expect a strong coupling between $\delta$, $C_{66}$ and $T_c$, which is, however, not observed. Spin-fluctuations, on the other hand, may be a candidate to mediate superconductivity, which is also suggested by their close correlation with $T_c$ under pressure \cite{Imai2009}. They appear not to be of the typical stripe-type nature and, thus, not strongly coupled to the structural distortion.
Inelastic neutron scattering would be useful in order to clarify the exact nature of the incipient magnetism in FeSe. 

We are grateful to R. M. Fernandes, U. Karahasanovic, S. Kasahara, H. Kontani, Y. Matsuda, J. Schmalian, and T. Shibauchi for stimulating discussions. A. B. also wishes to thank Kyoto University for the kind hospitality. This work has been supported
by the Japan-Germany Research Cooperative Program, KAKENHI from JSPS and Project No. 56393598 from DAAD.

%\bibliography{C:/Users/boehmer/Documents/reports/referencespnictidesall}

\begin{thebibliography}{40}
\bibitem [{not({\natexlab{b}})}]{notenemsus}%
  \BibitemOpen
  \href@noop {} {} \bibinfo {note}
  {$\lambda^2\chi_\varphi/C_{66,0}$ is obtained as in Ref. \cite{Boehmer2014},
  using the approximation $C_{66}/C_{66,0}\approx Y_{[110]}/Y_0$, where
  $Y_0/Y_{[110]}(250\textnormal{
  K})=1.3\left(1-s/\left(\exp(t/T)+1\right)\right)$ ($s=0.0421$ and $t=123.6$
  K) is a weakly temperature-dependent background from anharmonic phonons
  \cite{Varshni1970}.}\BibitemShut {Stop}%
\end{thebibliography}
%\bibliographystyle{apsrev4-1} 

\end{document}